\documentclass[10pt,journal]{IEEEtran}

\usepackage[latin1]{inputenc}
\usepackage[T1]{fontenc}

\usepackage{cite}
\usepackage[none]{hyphenat}
\usepackage{psfrag}
\usepackage[pdftex]{graphicx}
\usepackage{amsmath,amsfonts,amsbsy,amssymb}
\usepackage{mathrsfs}
\usepackage[ruled]{algorithm}
\usepackage{algorithmic}
\usepackage[nolist]{acronym}
\usepackage{float}

\usepackage{color}
\usepackage{soul}

\usepackage{tipa}

\usepackage[absolute,overlay]{textpos}
\usepackage{tikz}
\usetikzlibrary{topaths}
\usetikzlibrary{shapes,trees,decorations}
\usetikzlibrary{arrows}
\usetikzlibrary[shadows]
\usetikzlibrary{positioning}
\usetikzlibrary{matrix}
\definecolor{darkblue}{rgb}{.1,.1,.6}

\begin{document}

\title{Contention-based Geographic Forwarding Strategies for Wireless Sensors Networks}
\author{
	\IEEEauthorblockN{Carlos H.~M. de~Lima~\IEEEmembership{Member, IEEE}, Pedro. H. J. Nardelli~\IEEEmembership{Member, IEEE}, Hirley Alves~\IEEEmembership{Member, IEEE} and Matti Latva-aho~\IEEEmembership{Senior Member, IEEE}} 

	\thanks{C. H. M. de Lima is also with São Paulo State University (UNESP), São João da Boa Vista, Brazil. Email: carlos.lima@sjbv.unesp.br
		
		P. H. J. Nardelli, H. Alves and M. Latva-aho are with the Centre for Wireless Communications (CWC), University of Oulu, Finland. E-mail:\{nardelli, halves, matla\}@ee.oulu.fi. 
						
		The authors also would like to thank Academy of Finland, CNPq and Special Visiting Researcher fellowship CAPES 076/2012 from Brazil.}
}

\maketitle

\begin{acronym}[mmmmm]
	\acro{3GPP}[$3$GPP]{$3^\mathrm{rd}$ Generation Partnership Project}
	\acro{ABS}{Almost Blank Sub-frame}
	\acro{ADSL}{Asymmetric Digital Subscriber Line}
	\acro{ALBA-R}{Adaptive Load-Balanced Algorithm Rainbow}
	\acro{ALBA}{Adaptive Load-Balanced Algorithm}
	\acro{ALOHA}[ALOHA]{}
	\acro{APDL}{Average Packet Delivery Latency}
	\acro{AP}{Access Point}
	\acro{ASE}{Average Spectral Efficiency}
	\acro{BAP}{Blocked Access Protocol}
	\acro{BB}{Busy Burst}
	\acro{BC}{Broadcast Channel}
	\acro{BPP}{Binomial Point Process}
	\acro{BS}{Base Station}
	\acro{CAA}{Channel Access Algorithm}
	\acro{CAPEX}{Capital Expenditure}
	\acro{CAP}{Channel Access Protocol}
	\acro{CCDF}{Complementary Cumulative Distribution Function}
	\acro{CCI}{Co-Channel Interference}
	\acro{CDF}{Cumulative Distribution Function}
	\acro{CDMA}{Code Division Multiple Access}
	\acro{CDR}{Convex Lenses Decision Region}
	\acro{CF}{Characteristic Function}
	\acro{CGF}{Contention-based Geographic Forwarding}
	\acro{CM}{Coordination Mechanism}
	\acro{COMP}{Coordinated Multi-Point}
	\acro{CPICH}{Common Pilot Channel}
	\acro{CRA}{Conflict Resolution Algorithm}
	\acro{CRD}{Contention Resolution Delay}
	\acro{CRI}{Contention Resolution Interval}
	\acro{CRP}{Contention Resolution Protocol}
	\acro{CRS}{Cell-specific Reference Signal}
	\acro{CR}{Contention Resolution}
	\acro{CSMA/CA}{Carrier Sense Multiple Access with Collision Avoidance}
	\acro{CSMA/CD}{Carrier Sense Multiple Access with Collision Detection}
	\acro{CSMA}{Carrier Sense Multiple Access}
	\acro{CTM}{Capetanakis-Tsybakov-Mikhailov}
	\acro{CTS}{Clear To Send}
	\acro{D2D}{Device-To-Device}
	\acro{DAS}{Distributed Antenna System}
	\acro{DCF}{Distributed Coordination Function}
	\acro{DER}{Dynamic Exclusion Region}
	\acro{DHCP}{Dynamic Host Configuration Protoco}
	\acro{DL}{Downlink}
	\acro{DSL}{Digital Subscriber Line}
	\acro{E2E}{End-to-End}
	\acro{EDM}{Euclidean Distance Matrix}
	\acro{ES}{Evaluation Scenario}
	\acro{FAP}{Femto Access Point}
	\acro{FBS}{Femto Base Station}
	\acro{FD}{Full-Duplex}
	\acro{FDD}{Frequency Division Duplexing}
	\acro{FDM}{Frequency Division Multiplexing}
	\acro{FDR}{Forwarding Decision Region}
	\acro{FFR}{Fractional Frequency Reuse}
	\acro{FG}{Frequency Group}
	\acro{FPP}{First Passage Percolation}
	\acro{FUE}{Femto User Equipment}
	\acro{FU}{Femtocell User}
	\acro{GF}{Geographic Forwarding}
	\acro{GLIDER}{Gradient Landmark-Based Distributed Routing}
	\acro{GPSR}{Greedy Perimeter Stateless Routing}
	\acro{GeRaF}{Geographic Random Forwarding}
	\acro{HC}{Hard Core}
	\acro{HD}{Half-Duplex}
	\acro{HDR}[HDR]{High Data Rate}
	\acro{HII}{High Interference Indicator}
	\acro{HNB}{Home Node B}
	\acro{HN}[HetNet]{Heteronegeous Network}
	\acro{HOS}{Higher Order Statistics}
	\acro{HUE}{Home User Equipment}
	\acro{ICIC}{Inter-Cell Interference Coordination}
	\acro{IEEE}[IEEE]{}
	\acro{IMT}{International Mobile Telecommunications}
	\acro{IP}{Interference Profile}
	\acro{ITU}{International Telecommunication Union}
	\acro{KPI}{Key Performance Indicators}
	\acro{LN}{Log-Normal}
	\acro{LTE}{Long Term Evolution}
	\acro{LoS}{Line-of-Sight}
	\acro{M2M}{Machine-to-Machine}
	\acro{MACA}{Multiple Access with Collision Avoidance}
	\acro{MAC}{Medium Access Control}
	\acro{MBS}{Macro Base Station}
	\acro{MT}{Mellin Transform}
	\acro{MTC}{Machine Type Communication}
	\acro{MGF}{Moment Generating Function}
	\acro{MIMO}{Multiple-Input Multiple-Output}
	\acro{MPP}{Marked Point Process}
	\acroplural{MPP}[MPPs]{Marked Point Processes}
	\acro{MRC}{Maximum Ratio Combining}
	\acro{MS}{Mobile Station}
	\acro{MUE}{Macro User Equipment}
	\acro{MU}{Macrocell User}
	\acro{NB}{Node B}
	\acro{NLoS}{Non Line-of-Sight}
	\acro{NRT}{Non Real Time}
	\acro{OFDMA}{Orthogonal Frequency Division Multiple Access}
	\acro{OOP}{Object Oriented Programming}
	\acro{OPEX}{Operating Expenditure}
	\acro{OP}{Outage Probability}
	\acro{OS}{Order Statistic}
	\acro{PBS}{Pico Base Station}
	\acro{PCI}{Physical Cell Indicator}
	\acro{PC}{Power Control}
	\acro{PDF}{Probability Density Function}
	\acro{PDSR}{Packet Delivery Success Ratio}
	\acro{PGF}{Probability Generating Function}
	\acro{PMF}{Probability Mass Function}
	\acro{PPP}{Poisson Point Process}
	\acroplural{PPP}[PPPs]{Poisson Point Processes}
	\acro{PP}{Point Process}
	\acro{PRM}{Poisson Random Measure}
	\acro{PSS}{Primary Synchronization Channel}
	\acro{QoS}{Quality of Service}
	\acro{RAS}{Random Access System}
	\acro{RAT}{Radio Access Technology}
	\acro{RA}{Random Access}
	\acro{RCA}{Random Channel Access}
	\acro{RD}[R$\&$D]{Research $\&$ Development}
	\acro{REB}{Range Expansion Bias}
	\acro{RE}{Range Expansion}
	\acro{RF}{Radio Frequency}
	\acro{RIBF}{Regularized Incomplete Beta Function}
	\acro{RMA}{Random Multiple-Access}
	\acro{RNTP}{Relative Narrowband Transmit Power}
	\acro{RN}{Relay Node}
	\acro{RRM}{Radio Resource Management}
	\acro{RSA}{Relay Selection Algorithm}
	\acro{RSRP}{Reference Signal Received Power}
	\acro{RSSI}{Received Signal Strength Indicator}
	\acro{RSS}{Received Signal Strength}
	\acro{RS}{Relay Selection}
	\acro{RTS}{Request to Send}
	\acro{RT}{Real Time}
	\acro{RV}{Random Variable}
	\acro{SC}{Selection Combining}
	\acro{SDR}{Sectoral Decision Region}
	\acro{SF}{Sub-Frame}
	\acro{SG}{Stochastic Geometry}
	\acro{SI}{Self-Interference}
	\acro{SIC}{Successive Interference Cancellation}
	\acro{SINR}{Signal-to-Interference plus Noise Ratio}
	\acro{SIR}{Signal-to-Interference Ratio}
	\acro{SLN}{Shifted Log-Normal}
	\acro{SMP}{Semi-Markov Process}
	\acroplural{SMP}[SMPs]{Semi-Markov Processes}	
	\acro{SM}{State Machine}
	\acro{SNR}{Signal to Noise Ratio}
	\acro{SON}{Self-Organizing Network}
	\acro{SPP}{Spatial Poisson Process}
	\acroplural{SPP}[SPPs]{Spatial Poisson Processes}
	\acro{SSS}{Secondary Synchronization Channel}
	\acro{STA}{Splitting Tree Algorithm}
	\acro{TAS}{Transmit Antenna Selection}
	\acro{TCP}{Transmission Control Protocol}
	\acro{TC}{Transmission Capacity}
	\acro{TDD}{Time Division Duplexing}	
	\acro{TDMA}{Time Division Multiple Access}	
	\acro{TS}{Terminal Station}
	\acro{TTI}{Transmission Time Interval}
	\acro{TTT}{Time To Trigger}
	\acro{UDM}{Unit Disk Model}
	\acro{UD}{Unit Disk}
	\acro{UE}{User Equipment}
	\acro{ULUTRANSIM}[UL UTRANSim]{R6 Uplink UTRAN Simulator}
	\acro{UL}{Uplink}
	\acro{UML}{Unified Modeling Language}
	\acro{UoI}{User of Interest}
	\acro{UMTS}{Universal Mobile Telecommunications System}
	\acro{WCDMA}{Wideband Code Division Multiple Access}
	\acro{WSN}{Wireless Sensor Network}
	\acro{iid}[\textup{i.i.d.}]{independent and identically distributed}
\end{acronym}

\begin{abstract}
	This paper studies combined relay selection and opportunistic geographic
	routing strategies for autonomous wireless sensor networks where
	transmissions occur over multiple hops.	
	The proposed solution is built upon three constituent parts: {($i$)}
	relay selection algorithm, {($ii$)} contention resolution mechanism, and
	{($iii$)} geographic forwarding strategy.
	Using probability generating function and spatial point process as the
	theoretic background, we propose an auction-based algorithm for selecting
	the relay node that relies on the network topology as
	side-information.
	Our results show that geographic solutions that iteratively exploit the
	local knowledge of the topology to ponder the algorithm operation
	outperforms established random approaches.
\end{abstract}


\acresetall

\section{Introduction}
\label{SEC:INTRO}

Lately, wireless communications are becoming a key technology for
autonomous networks by providing means to deploy a wide range of emerging
applications, such as smart houses, smart factories and networked cars.
Wireless sensor networks are attractive for many reasons: low implementation
and maintenance costs, flexible (physical) topology, as well as scalability
\cite{ART:Liu-SJ15}.
Additionally, it enable information exchange between autonomous devices without
any (direct) human intervention \cite{ART:Liu-SJ15, ART:HAN-CM14,
Stojmenovic2014}.
%
%
%
%
%
However, {a major challenge for wireless sensor} networks is to cope with the inherent requirements of {such} application.
For instance, in {an} industrial environment these requirements are often
more stringent than home environments due to the presence of highly reflective
materials and additional interference from the machinery, which limits
performance due to the increase interference profile \cite{Stenumgaard2013}.
Then, {legacy} point-to-point or point-to-multipoint {protocols}, which are well-established concepts in many wireless systems, does not seem to be suitable for industrial \cite{Stenumgaard2013}.

{In this context}, the use of short hops to form a
wireless multi-hop link appear to be a simple, while efficient, solution for
many applications from home to industrial environments. 
The literature of ad hoc networks provides several insights on how to analyze
and build multi-hop systems.
For instance, \cite{baccelli2010time, srinivasa2014combining} study the
formation and maintenance of multi-hop connections in large-scale ad hoc
networks.
As a result of nodes' mobility, network dynamics and channel impairments, the wireless links undergo great fluctuation on their availability
and quality.
Specifically in \cite{baccelli2010time}, Baccelli {\textit{et al.}}
emphasize that opportunistic routing schemes, which dynamically form multi-hop
links by selecting the most suitable relay at any slot and at any hop,
outperform other routing mechanisms in such distributed scenarios.
In {fact}, there are no fixed routers: any node in the
network should be capable of relaying packets.

{Additionally, as} control and payload information share the same
pool of available resources, the route management should avoid excessive
overhead.
\ac{QoS} requirements must be also satisfied when optimizing the utilization of
the network resources.
When dealing with multi-hopping, the design decision of having a route over
many short hops or over few long hops (i.e hopping strategy) is critical
\cite{srinivasa2014combining,Nardelli2012}.
However, regardless the strategy used, there still exists a negotiation period
between the nodes to decide which is the most suitable relay.

When evaluating distributed geographic routing strategies, the most used
approach is to assess the \ac{CGF} solutions in terms of the progress they
provide towards the destination.
For instance in \cite{ART:CHEN-TVT07}, the authors addressed the problem of
defining the forwarding regions, determining their impact on the system
performance.
This provides clear guidelines for designing the geographic routing
protocols.
Then, \cite{PROC:LIMA-WCNC08} proposes a greed forwarding cluster-based
algorithm which is resilient to topological variations due to network dynamics,
which is shown to enhance system performance with low latency.
In \cite{PROC:LIMA-ITW09, PROC:LIMA-WPNC09} an initial assessment of
contention-based relay selection strategies is performed. 

This paper extends those results while focusing on both the packet expected
forward distance and the cost of finding a relay at each hop.
The analysis of the geographic routing strategies is conducted by assessing the
relay selection algorithms, the contention resolution mechanisms, and the
geographic forwarding schemes.
Specifically, the cost of selecting the next hop relay is characterized in
terms of the distribution of the time necessary to resolve the contention, also
known as the relay election process.
The expected progress obtained at each hop is determined using analytic tools
of stochastic geometry.

Then, we introduce two distinct \ac{CGF} schemes combining geographic
forwarding designs and \acp{RSA}.
The first one combines \ac{SDR} with \ac{STA}-based\footnote{It is noteworthy
	that a comprehensive review of hierarchical routing protocols is provided
	in \cite{ART:Liu-SJ15}.} \ac{RSA}, while the second scheme jointly employs
	\ac{CDR} and  auction-based \ac{RSA}.

{Herein, we introduce an analytical framework rather than using a network
level simulation-based evaluation as presented in} \cite{PROC:LIMA-WCNC08}.
{Moreover, we extend the analysis provided in} \cite{PROC:LIMA-WCNC08,
PROC:LIMA-ITW09, PROC:LIMA-WPNC09}.
{Our analytical framework is employed to first characterize the} \ac{CRI}
{of the selection algorithms, and then compare the performance of these
	different solutions.
Finally, we assess how the network performs in terms of the packet expected
forward distance and the cost of finding a relay at each hop.}
{Hence, our} contributions are summarized as follows:
%
\begin{itemize}
	\item characterization of the statistics of the \ac{CRI} length for the contention-based \acp{RSA};
	\item evaluation framework to assess the achievable progress attained by different geographic forwarding regions; 
	\item new auction-based relay selection scheme for \ac{RMA} networks that recursively adapt the forwarding regions; and 
	\item unifying framework to jointly analyze the achievable
	progress and the negotiation overhead.
\end{itemize}

The reminder of this paper is organized as follows: Section \ref{SEC:SYS_MODEL} introduces the system model. In Section \ref{SEC:RSA} describes the relay selection procedures, while Section \ref{SEC:ROUTING} presents the geographic forwarding strategies. Next, Section \ref{SEC:RESULTS} provides an comprehensive set of numerical results and discussion. Finally, Section \ref{SEC:CONCLUSIONS} concludes the paper. 

\section{Problem Description}
\label{SEC:SYS_MODEL}

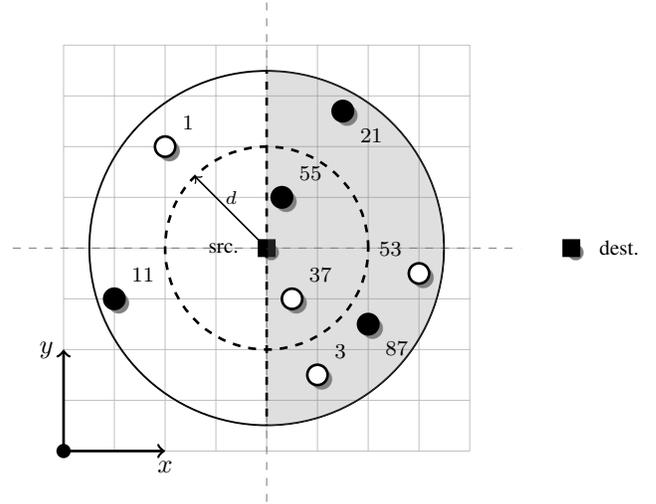
\begin{figure}[!t]
	\centering
	\begin{tikzpicture}[
	scale = 1.35,
	active/.style = {fill = black, line width = 1pt, circular drop shadow},
	idle/.style = {fill = white, line width = 1pt, circular drop shadow}
	]
	\draw[step = 5mm, help lines, opacity = .5] (-2, -2) grid (2, 2);
	\fill (-2, -2) circle (2pt);
	\draw[<->, line width = 1pt] (-1, -2) node[below] {$x$} -| (-2, -1)
	node[left] {$y$};
	\draw[dashed, gray] (0, -2.5) -- (0, 2.5);
	\draw[dashed, gray] (-2.5, 0) -- (2.5, 0);
	\node[left = .25cm] (src) at (0, 0) {\footnotesize src.};
	\filldraw[active] (-.075, -.075) rectangle +(.15,.15);
	\node[right = .25cm] (dest) at (3, 0) {\footnotesize dest.};
	\filldraw[active] (2.925, -0.075) rectangle +(.15,.15);
	\clip[draw] (0, 0) circle (1.75cm);
	\filldraw[fill = none, line width = 1pt] (0, 0) circle (1.75cm);
	\draw[dashed, line width = 1pt, fill = gray, fill opacity = .25]
	(0,0) -- (0,2) -- (1.75,2) -- (1.75,-2) -- (0,-2) -- cycle;
	\node[below right = .15cm] (a) at (.75, 1.35) {\footnotesize $21$};
	\filldraw[active] (.75, 1.35) circle (.1cm);
	\node[above left = .15cm] (b) at (1.5, -.25) {\footnotesize $53$};
	\filldraw[idle] (1.5, -.25) circle (.1cm);
	\node[below right = .15cm] (c) at (1, -.75) {\footnotesize $87$};
	\filldraw[active] (1, -.75) circle (.1cm);
	\node[above right = .15cm] (d) at (.5, -1.25) {\footnotesize $3$};
	\filldraw[idle] (.5, -1.25) circle (.1cm);
	\node[above right = .15cm] (e) at (.25, -.5) {\footnotesize $37$};
	\filldraw[idle] (.25, -.5) circle (.1cm);
	\node[above right = .15cm] (f) at (.15, .5) {\footnotesize $55$};
	\filldraw[active] (.15, .5) circle (.1cm);
	\node[above right = .15cm] (g) at (-1, 1) {\footnotesize $1$};
	\filldraw[idle] (-1, 1) circle (.1cm);
	\node[above right = .15cm] (h) at (-1.5, -.5) {\footnotesize $11$};
	\filldraw[active] (-1.5, -.5) circle (.1cm);
	\filldraw[fill = none, line width = 1pt, dashed] (0,0) circle(1cm);
	\node (a) at (-.35,.5) {\scriptsize $d$};
	\draw[->, line width=.6pt] (0,0) +(0:0cm) -- +(135:1cm);
	\end{tikzpicture}
	\caption{Illustration of the sectoral decision region with $Q=2$ splitting
		groups and angular aperture of $180$ degrees. Dashed lines defined the
		forwarding region (shaded in light gray), while black circles
		identify awake nodes and white circles identify asleep nodes. All awake
		neighbors within the shaded region are eligible relays.}
	\label{FIG:CONFIG_SDR}
\end{figure}

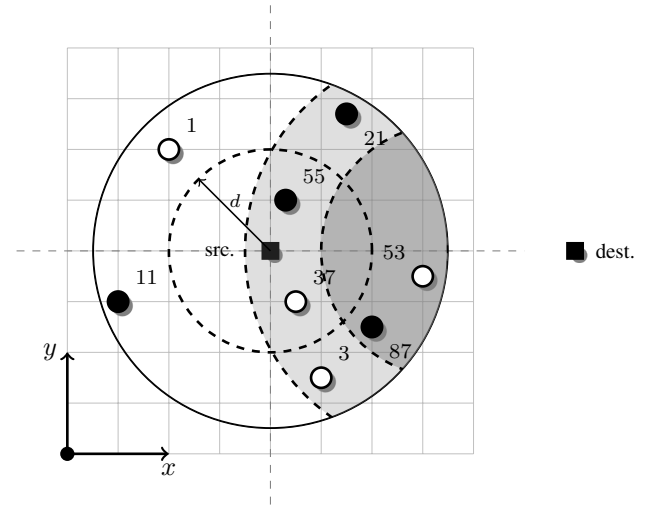
\begin{figure}[!t]
	\centering
	\begin{tikzpicture}[
	scale = 1.35,
	active/.style = {fill = black, line width = 1pt, circular drop shadow},
	idle/.style = {fill = white, line width = 1pt, circular drop shadow}
	]
	\draw[step = 5mm, help lines, opacity = .5] (-2, -2) grid (2, 2);
	\fill (-2,-2) circle (2pt);
	\draw[<->, line width = 1pt] (-1, -2) node[below] {$x$} -| (-2,-1)
	node[left] {$y$};
	\draw[dashed, gray] (0, -2.5) -- (0, 2.5);
	\draw[dashed, gray] (-2.5, 0) -- (2.5, 0);
	\node[left = .35cm] (src) at (0, 0) {\footnotesize src.};
	\filldraw[active] (-.075, -.075) rectangle +(.15, .15);
	\node[right = .15cm] (dest) at (3, 0) {\footnotesize dest.};
	\filldraw[active] (2.925, -0.075) rectangle +(.15, .15);
	\clip[draw] (0, 0) circle (1.75cm);
	\filldraw[fill = none, line width = 1pt] (0, 0) circle (1.75cm);
	\draw[dashed, line width = 1pt, fill = gray, fill opacity = .25] (1.75, 2)
	arc (90:270:2cm) -- (1.75, 2) -- cycle;
	\draw[dashed, line width=1pt, fill=gray, fill opacity = .45] (1.75, 1.25)
	arc (90:270:1.25cm) -- (1.75, 1.25) -- cycle;
	\node[below right = .15cm] (a) at (.75, 1.35) {\footnotesize $21$};
	\filldraw[active] (.75, 1.35) circle (.1cm);
	\node[above left = .15cm] (b) at (1.5, -.25) {\footnotesize $53$};
	\filldraw[idle] (1.5, -.25) circle (.1cm);
	\node[below right = .15cm] (c) at (1, -.75) {\footnotesize $87$};
	\filldraw[active] (1, -.75) circle (.1cm);
	\node[above right = .15cm] (d) at (.5, -1.25) {\footnotesize $3$};
	\filldraw[idle] (.5, -1.25) circle (.1cm);
	\node[above right = .15cm] (e) at (.25, -.5) {\footnotesize $37$};
	\filldraw[idle] (.25, -.5) circle (.1cm);
	\node[above right = .15cm] (f) at (.15, .5) {\footnotesize $55$};
	\filldraw[active] (.15, .5) circle (.1cm);
	\node[above right = .15cm] (g) at (-1, 1) {\footnotesize $1$};
	\filldraw[idle] (-1, 1) circle (.1cm);
	\node[above right = .15cm] (h) at (-1.5, -.5) {\footnotesize $11$};
	\filldraw[active] (-1.5, -.5) circle (.1cm);
	\filldraw[fill = none, line width = 1pt, dashed] (0,0) circle(1cm);
	\node (a) at (-.35,.5) {\scriptsize $d$};
	\draw[->, line width=.6pt] (0,0) +(0:0cm) -- +(135:1cm);
	\end{tikzpicture}
	\caption{Illustration of the second iteration of the convex lenses decision
		region with $Q=2$ regions.  Dashed lines defined the forwarding
		region (shaded regions in light gray), while black circles identify awake
		nodes and white circles identify asleep nodes. All awake neighbors dwelling
		within the shaded region are eligible relays.}
	\label{FIG:CONFIG_CDR}
\end{figure}

 
We assume that potential relay nodes are randomly distributed over the network
area.
Figs. \ref{FIG:CONFIG_SDR} and \ref{FIG:CONFIG_CDR} illustrate the deployment
model and the geographic forwarding regions, namely, sectoral
and convex lenses decision region, respectively.
The distribution of the number of neighbors within source's transmission range
corresponds to a general $2$-dimensional binomial point process: the number of
points within a given region is fixed while their positions are uniformly
distributed.

A simplified connectivity model based on the unit disk graph is employed.
In this random connectivity model, the Euclidean distances between nodes
determine their connectivity.
Discs of equal diameter form a graph in which any two vertices are connected by an edge whenever one disc contains the center of
the other.
In this case, awake neighbors within the source's radio range
are considered eligible relays.
Yet nodes dwelling in the source's transmission range
successfully receive packets and the only cause of errors are packet
collisions.
In what follows, it is assumed that source nodes are not affected by
the hidden and exposed problems, since we are specifically addressing the
iterations among potential relays.

Nodes operate synchronously and packets are transmitted on a slot-by-slot
basis.
Nodes use the shared medium (air interface) by means of a contention-based
random multi-access scheme.
A two state error-less channel model is used, where the air interface is either
busy or idle.
Assume a destructive interference amongst concurrent transmissions in
which the only source of packet loss is collision.
Every network node operates in half-duplex fashion, and therefore can both transmit and receive packets, though not simultaneously.

Time-slotted collision-type channel with binary feedback (either collision or
no collision) and gated channel access are employed.
By the end of each transmission slot nodes are immediately and errorlessly
aware of the feedback.
The \acp{CAA} specifies when packets may join the contention resolution
transactions.
New packets that appear during the resolution of the current conflict are
buffered, i.e. the access to the channel is blocked to all that did not take
part in the colliding slot originating the \ac{CRI} that is afoot.
In other words, any potential relay that wakes up during an ongoing \ac{CRI}
does not interferer on the ongoing transactions.

\vspace{-0.3cm}
\section{Contention-based \acl{RSA}}
\label{SEC:RSA}

Two distinct contention-based relay selection mechanisms are analyzed: ($i$)
a totally random solution based solely on the standard
splitting tree algorithm for performing \ac{RMA} communications
\cite{ART:YU-ITIT07}; and ($ii$) an
auction-based approach exploiting the local knowledge of network topology to
avoid collisions and, whenever necessary, to shorten the contention resolution
period \cite{PROC:LIMA-ITW09}.

The contention resolution algorithms are classified in terms of the endured
\ac{CRI} length necessary to select the most suitable relay among all the
eligible neighbors.
It is worth to emphasize that only the time necessary to resolve the
contentions is considered to analyze the negotiation cost of the \acp{CRA}.
Neither the queuing time nor the (re)transmissions propagation time of the data
payload are taken into consideration for the overhead assessment.

To characterize the cost of selecting a relay at each hop, \acp{PGF} are used to represent the \ac{PMF} of the \ac{CRI} length by means of power series with non-negative coefficients.
The \ac{PMF} is thereby recovered by numerically inverting the corresponding
\ac{PGF} using an approximation based on the Fourier series method.
%
The distribution of the \ac{CRI} length, namely $L_N$, is approximated as:
\begin{eqnarray}
\label{EQ:INV_PGF_APPROX} 
\widetilde{\Pr}\left\{L_N\negthinspace = k \right\}\negthinspace = \frac{1}{ 2 k {r^k} }
	\sum\limits_{j = 1}^{2 k}{ { (-1) ^j} \Re\left[ {G}_{N}\left( r
	{ e^{ \frac{ \pi j i }{ k } } } \right) \right] },
\end{eqnarray}
{\noindent where radius $r$ (the radius of convergence).}

\subsection{ \ac{PGF} of the \ac{STA}-based \ac{RSA}}
\label{SEC:STA}

The splitting tree algorithm constitutes the operational underpinning of the
\ac{STA}-based solution, which is tailored to \ac{RMA} communications
\cite{ART:YU-ITIT07}.
The \ac{CRA} works with the \ac{CAA}, which in its turn
dictates when new packets may join the transactions still in progress.
A gated channel access algorithm, known as \ac{BAP}, is  used to
control the ingress of new packets in the contention.
Meaning that no new contending relay is allowed in the contention taking place once the resolution of the conflict has been already initiated.

According to this on-demand \ac{MAC} mechanism, the source node always initiates the relay selection transactions by issuing a \ac{RTS} packet.
The neighbors that listened to the source's requisition split
themselves randomly and independently based on the common probability that dictates the likelihood of accessing the shared channel -- totally random approach.
If no suitable relay is found in a given relay selection
interaction, the source node backs off.
After a predefined interval, it restarts the relay selection procedure
addressing (hopefully) new players.

Since the \ac{STA}-based scheme is a random approach to select relays, the source node can only identify the most suitable relay after collecting the forwarding information from all the eligible relays.
The performance of the \ac{STA}-based \ac{RSA} is addressed by means of
computational simulations in \cite{PROC:LIMA-WCNC08}.
Fig. \ref{FIG:CONFIG_SDR} illustrates a snapshot of the sectoral decision region that is used in conjunction with the \ac{STA}-based solution.

The \ac{PGF} of ${L}_{N}$ is the conditional \ac{CRI}
length when $N$ nodes initially collide.
The conditional \ac{CRI} length considering a $Q$-sided fair coin is: ${L}_{N}=1$ for $N\in\{0,1\}$, or ${L}_{N}=1+ \sum_{j = 1}^{Q} { L_{I_j} }$ when $N\geq2$.
Note that the computation of the statistics of the ${L}_{N}$ involves the summation of multiple \acp{RV} each of them corresponding to a particular splitting group (or even subsets thereof).

For example, the \ac{PMF} of the sum of only two independent
discrete \acp{RV} $X$ and $Y$ is given by the
convolution of their corresponding \acp{PMF}, $i.e.$ $\displaystyle p_{X+Y} = p_X * p_Y$.
One can see that computing the statistics of the
${L}_{N}$ (total \ac{CRI} length) by convolving all subsets would be a laborious task.
For that reason, \acp{PGF} are conveniently used herein so as to derive the distribution of the \ac{CRI} length.

The \ac{PGF} of ${L}_{N}$ for the \ac{STA}-based approach is expressed as follows:
\begin{equation}
	\label{EQ:PGF_CRI_DEF}
	{G}_{N}(z) = \sum\limits_{k = 0}^{\infty} { \Pr\left\{ L_N = k
	\right\} z^k } = \textup{E} \left\{ z^{L_N} \right\},
\end{equation}
{\noindent where ${L}_{N}$ is a discrete \ac{RV} assuming non-negative
integer values representing the contention resolution intervals.}
Note that it follows from the definition of the \ac{PGF} of the
${L}_{N}$ that ${G}_{0}(z)={G}_{1}(z)=z$. 

The expectation of \eqref{EQ:PGF_CRI_DEF} is then computed as
\begin{align}
	\label{EQ:EXPECTATION_CRI_PGF}
	&\textup{E} \left\{ z^{L_N} \right\} = \textup{E} \left\{ \textup{E} \left\{ z^L \vert { I_1,..., I_Q } \right\} \vert N \right\},
\end{align}
{\noindent which yields}
\begin{equation}\label{EQ:PGF_MULTINOMIAL}
	{G}_{N}(z) = z \sum\limits_{ i_1, \dots, i_Q }^{ N }{ \binom{N}{ i_1,..., i_Q } \prod_{j=1}^Q { {Q}_{ L_{I_1} }(z) {P_j}^{i_j} } },
\end{equation}
{\noindent where ${P_j}^{i_j}$ is the probability of $i_j$ nodes flip the $j$
side of the $Q$-sided fair coin.}

Note that the summation iterates over all possible combinations of the
multinomial splitting groups $i_1,..., i_Q$.
Next the \ac{CRI} length is particularized for the binary tree
configuration alone (two splitting groups).
Thus, $B_{N,i} = \binom{N}{i} { {p}^{i} }{ {\left( 1 - p \right)}^{N - i} }$  yields the probability that exactly $i$ nodes toss $0$ (first splitting group), and then transmit in the very next frame after the collision, 
where $p$ corresponds to the probability of tossing $0$ when using
the unbiased binary coin for each choice.
Finally, the \ac{PGF} of the \ac{CRI} length for the \ac{STA}-based relay
selection scheme is then given by,
\begin{align}\label{EQ:PGF_BINARY_TREE}
	{G}_{N}(z) &= z \sum\limits_{i = 0}^{N}{ B_{N,i} {G}_{i}(z) {G}_{N - i}(z) },
\end{align}
{\noindent where ${G}_{i}(z)$ addresses the collision among $i$ nodes that
flipped $0$ (1st subset), and ${G}_{N - i}(z)$ corresponds to the additional slots to resolve the collision among $N-i$ nodes that flipped $1$ (2nd subset).}

\begin{figure}[!t]
	\centering
	\psfrag{titlea}[Bc][Bc][1][0]{PMF of the CRI length}
	\psfrag{titleb}[Bc][Bc][1][0]{STA-based Relay Selection Algorithm}
	\psfrag{ylabel}[Bc][Bc][1][0]{$\widetilde{\Pr} \left\{ L_N = k \right\}$ (\%)}
	\psfrag{xlabel}[Bc][Bc][1][0]{$k$ (slots)}
	
	\psfrag{N=2, Analysis}[Bl][Bl][1][0]{\hspace{-0em}$N=2$, Analysis}
	\psfrag{N=2, Simulation}[Bl][Bl][1][0]{\hspace{-0em}$N=2$, Simulation}
	\psfrag{N=3, Analysis}[Bl][Bl][1][0]{\hspace{-0em}$N=3$, Analysis}
	\psfrag{N=3, Simulation}[Bl][Bl][1][0]{\hspace{-0em}$N=3$, Simulation}
	\psfrag{N=4, Analysis}[Bl][Bl][1][0]{\hspace{-0em}$N=4$, Analysis}
	\psfrag{N=4, Simulation}[Bl][Bl][1][0]{\hspace{-0em}$N=4$, Simulation}
	\psfrag{N=5, Analysis}[Bl][Bl][1][0]{\hspace{-0em}$N=5$, Analysis}
	\psfrag{N=5, Simulation}[Bl][Bl][1][0]{\hspace{-0em}$N=5$, Simulation}
	
	\includegraphics[width=1\columnwidth]{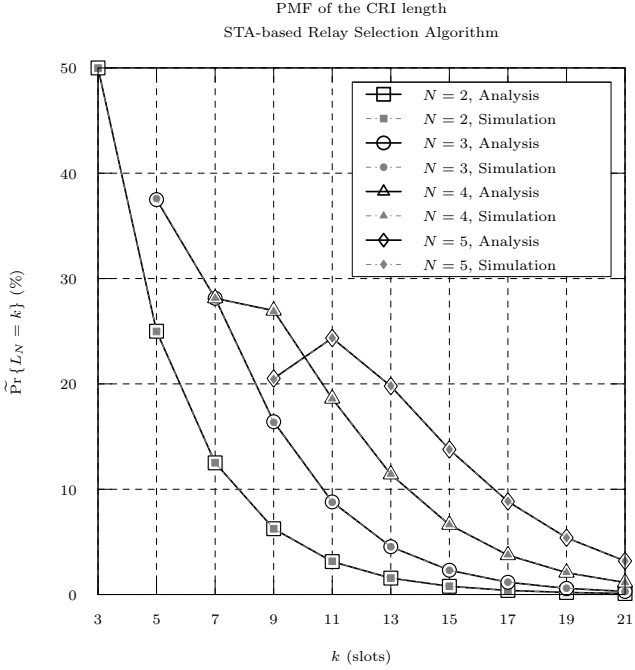}
	\vspace{-5mm}
	\caption{\ac{PMF} of the \ac{CRI} length for the \ac{STA}-based \ac{RSA}.
		$N$ is the multiplicity of the conflict and corresponds to the number of
		relays that initially collide. We consider $Q=2$ 
		groups.}
	\label{FIG:STA_CRI}
\end{figure}

Fig. \ref{FIG:STA_CRI} presents the \ac{PMF} of the \ac{CRI} length when the \ac{STA}-based \ac{CRA} is employed.
The numbers of nodes identify the number of candidate relays
that initially collide.
The distribution of the \ac{CRI} length is generated for an increasing number
of contending relays.
When using the totally random approach, the \ac{CRI} significantly lengthens
with the initial number of colliding relays.
In fact, the resolution of the conflict may linger too much time before the
contention is resolved and the next-hop relay is elected.
As a consequence, the achievable data rate per node is compromised.

It is worth noting that when randomly interacting through the relay election
process, nodes may expose themselves to much higher channel contention and
consequently squander the already limited radio resources.
To reduce the conflict over the air interface, potential relays may
ponder their participation in the election process beforehand by dynamically
evaluating the characteristics affecting the transmission, such as local
network topology, network dynamics and channel impairments.

\vspace{-0.45cm}
\subsection{\ac{PGF} of the Auction-based \ac{RSA}}
\label{SEC:AUCTION}

The auction-based \ac{RSA} iteratively exploits location information to solve
conflicts and then shortens the resolution
interval by pruning the binary conflict resolution tree.
In \cite{PROC:LIMA-WCNC08}, Dutch auctions\footnote{Dutch auctions are extremely convenient to sell goods -- assignment of network resources -- quickly.
The reasons are two-fold, the auction ends with the very first bid and the
auctioneer may appropriately set the decreasing rate of the artifact value
(depreciation rate) aiming at quickening the auction.} are proposed as an effective alternative to address the relay selection process in conjunction with \ac{RMA} methods, where the source is the ``auctioneer'' and potential relays are the ``bidders''.
The price is then derived from the separation between source, relays and
destination (similar to the simple greedy forwarding)
\cite{ART:JI-JSAC08}.

Fig. \ref{FIG:CONFIG_CDR} illustrates the computation of the forwarding
regions regarding two splitting groups.
In the first round of the auction, the candidate relays are divided into two groups, namely $\left\{ \left\{ 87, 21, 55 \right\}, \left\{ 11 \right\} \right\}$.
Since nodes $\left\{ 87, 21, 55 \right\}$ reply at the same time slot, a
collision occurs.
The source detects the collision in the first slot and recomputes the
forwarding regions accordingly. 
Whenever the \ac{BAP} is considered, the contending nodes can also recompute the forwarding regions independently by themselves.
Node $\left\{ 11 \right\}$ also detects the collision and, as nodes of higher priority have already replied, just drops out.
Thereafter, relays in the first colliding area are reordered in the sequence
$\left\{ \left\{ 87 \right\}, \left\{ 21, 55 \right\} \right\}$.
Finally, node $\left\{ 87 \right\}$ replies and the auction finishes.

According to the auction-based \ac{RSA}, potential relays that have not
replied in the previous slot but detected a collision drop out of the
ongoing transaction.
This intrinsic ``tree pruning'' means that whenever the first
subset visited after a collision leads to another collision, the second subset is dropped.
For $\displaystyle i > 1$, instead of the \ac{CRI} having full
length $\displaystyle B_{N,i} {G}_{i}(z) {G}_{N - i}(z) $, the tree pruning procedure leads to a shorter length $\displaystyle B_{N,i} {G}_{i}(z)$ \cite{mohamed2010dynamic}.

Note that the derivation of the \ac{PGF} of the \ac{CRI} length is
conditioned on the multiplicity of the initial set of colliding nodes.
We can then express the \ac{PGF} of the auction-based \ac{RSA} as
follows:
\begin{eqnarray}
	\label{EQ:PGF_AUCTION}
	{G}_{N}(z)\negthinspace=\negthinspace z^2 B_{N,0}
	{G}_{N}(z)\negthinspace+\negthinspace z^2 B_{N,1} \negthinspace+\negthinspace z
	\sum\limits_{i =2}^{N}{\negthinspace B_{N,i} {G}_{i}(z)},
\end{eqnarray}
{\noindent where the first term accounts for case when no reply is issued in the first slot, and the second term addresses the case when there is only one eligible node in the first decision region (first slot).}

Whenever nodes involved in a contention listen to an idle slot just after the colliding slot they do not need to undergo a collision in the subsequent slot.
The auction-based solution can be refined in a way that nodes
may split themselves into the decision regions prior to any indication of
collision whereby the collision avoidance mechanism is characterized.
This procedure leads to skipping one level of
the binary tree, and is expressed by dropping one
slot of the first term of \eqref{EQ:PGF_AUCTION}, which then becomes
\begin{eqnarray}
\label{EQ:PGF_AUCTION_SKIP}
{G}_{N}(z)\negthinspace =\negthinspace z B_{N,0} {G}_{N}(z)\negthinspace + z^2 B_{N,1} \negthinspace+ z \sum\limits_{i =2}^{N}{\negthinspace B_{N,i} {G}_{i}(z) }.
\end{eqnarray}

\begin{figure}[!t]
	\centering
	
	\psfrag{titlea}[Bc][Bc][1][0]{PMF of the CRI length}
	\psfrag{titleb}[Bc][Bc][1][0]{Auction-based Relay Selection Algorithm}
	\psfrag{ylabel}[Bc][Bc][1][0]{$\widetilde{\Pr} \left\{ L_N = k \right\}$ (\%)}
	\psfrag{xlabel}[Bc][Bc][1][0]{$k$ (slots)}
	
	\psfrag{N=2, Analysis}[Bl][Bl][1][0]{\hspace{-0em}$N=2$, Analysis}
	\psfrag{N=2, Simulation}[Bl][Bl][1][0]{\hspace{-0em}$N=2$, Simulation}
	\psfrag{N=3, Analysis}[Bl][Bl][1][0]{\hspace{-0em}$N=3$, Analysis}
	\psfrag{N=3, Simulation}[Bl][Bl][1][0]{\hspace{-0em}$N=3$, Simulation}
	\psfrag{N=4, Analysis}[Bl][Bl][1][0]{\hspace{-0em}$N=4$, Analysis}
	\psfrag{N=4, Simulation}[Bl][Bl][1][0]{\hspace{-0em}$N=4$, Simulation}
	\psfrag{N=5, Analysis}[Bl][Bl][1][0]{\hspace{-0em}$N=5$, Analysis}
	\psfrag{N=5, Simulation}[Bl][Bl][1][0]{\hspace{-0em}$N=5$, Simulation}
	
	\includegraphics[width=1\columnwidth]{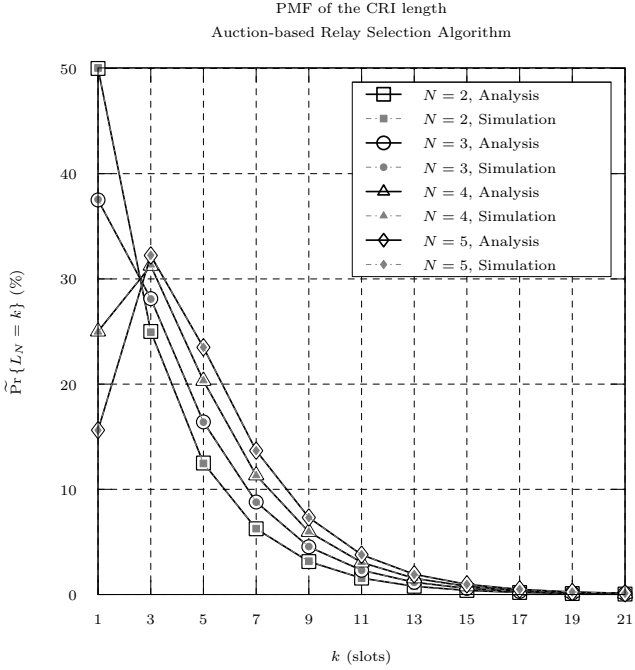}
	\vspace{-5mm}
	\caption{\ac{PMF} of the \ac{CRI} length for the auction-based \ac{RSA}.
		$N$ is the multiplicity of the conflict and corresponds to the 
		candidate relays that initially collide. We consider $Q=2$ groups.}
	\label{FIG:AUC_CRI}
\end{figure}

For the auction-based \ac{RSA}, the \ac{PMF} of the \acp{CRI} length is
presented in Fig. \ref{FIG:AUC_CRI}.
The impact of the initial number of colliding nodes on the duration of the
resolution is still observed, though in much lesser extent.
By using the location information, eligible nodes can
independently split themselves into priority groups quickening the selection transactions in a distributed manner.
The location awareness improves the
contention resolution capability of the auction-based alternative, mainly when  the initial number of colliding nodes are high.

From Figs. \ref{FIG:STA_CRI} and \ref{FIG:AUC_CRI}, regarding the curves of four eligible relays alone, the resolution probability for nine slots is nearly $28\,\%$ for the \ac{STA}-based procedure, whereas the auction-based one is approximately $8\,\%$ only.

\section{\acl{GF} Strategies}
\label{SEC:ROUTING}

We address the geographic forwarding component of the \ac{CGF} strategies by characterizing the impact of distinct decision regions on the expected progress of a packet towards its final destination at each single hop.
The \ac{CGF} strategies are a common topic in the literature when assessing position-centric network routing solutions \cite{ART:CHEN-TVT07,Ruhrup2010}.

The geographic forwarding strategies make use of the knowledge of the network topology, globally or locally, to route packets along multi-hop links.
Relays are selected in a hop basis depending on their relative positions  in relation to the source: the node that provides the longest progress towards the destination is chosen.
%

\begin{figure}[!t]
	\centering
	
	\psfrag{titlea}[Bc][Bc][1][0]{Sectoral Decision Region}
	\psfrag{titleb}[Bc][Bc][1][0]{Distance to the $n$-th nearest neighbor}
	\psfrag{ylabel}[Bc][Bc][1][0]{$f_{D_n}\left(d \right)$}
	\psfrag{xlabel}[Bc][Bc][1][0]{$d$ (m)}
	
	\psfrag{Analysis}[Bl][Bl][1][0]{\hspace{-0em}Analysis}
	\psfrag{Simulation}[Bl][Bl][1][0]{\hspace{-0em}Simulation}
	\psfrag{1st Relay}[Bl][Bl][1][0]{\hspace{-0em}$1^\textrm{st}$ Neighbor}
	\psfrag{2nd Relay}[Bl][Bl][1][0]{\hspace{-0em}$2^\textrm{nd}$ Neighbor}
	\psfrag{3rd Relay}[Bl][Bl][1][0]{\hspace{-0em}$3^\textrm{rd}$ Neighbor}
	\psfrag{4th Relay}[Bl][Bl][1][0]{\hspace{-0em}$4^\textrm{th}$ Neighbor}
	\psfrag{5th Relay}[Bl][Bl][1][0]{\hspace{-0em}$5^\textrm{th}$ Neighbor}
	
	\includegraphics[width=1\columnwidth]{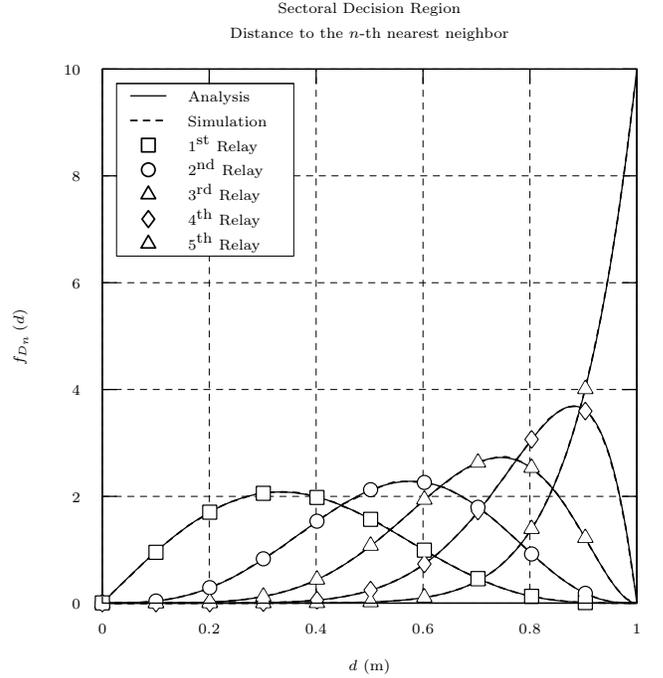}
	\vspace{-5mm}
	\caption{\ac{PDF} of the distance from a reference point
		to the $n$-th neighbor when using the \ac{SDR}
		design.  The transmission range is $R=1$m and the number of 	candidate relays is $N=5$.}
	\label{FIG:SDR_PDF}
\end{figure}
\begin{figure}[!t]
	\centering
	
	\psfrag{titlea}[Bc][Bc][1][0]{Convex Lenses Decision Region}
	\psfrag{titleb}[Bc][Bc][1][0]{Distance to the $n$-th nearest neighbor}
	\psfrag{ylabel}[Bc][Bc][1][0]{$f_{D_n}\left(d \right)$}
	\psfrag{xlabel}[Bc][Bc][1][0]{$d$ (m)}
	
	\psfrag{Analysis}[Bl][Bl][1][0]{\hspace{-0em}Analysis}
	\psfrag{Simulation}[Bl][Bl][1][0]{\hspace{-0em}Simulation}
	\psfrag{1st Relay}[Bl][Bl][1][0]{\hspace{-0em}$1^\textrm{st}$ Neighbor}
	\psfrag{2nd Relay}[Bl][Bl][1][0]{\hspace{-0em}$2^\textrm{nd}$ Neighbor}
	\psfrag{3rd Relay}[Bl][Bl][1][0]{\hspace{-0em}$3^\textrm{rd}$ Neighbor}
	\psfrag{4th Relay}[Bl][Bl][1][0]{\hspace{-0em}$4^\textrm{th}$ Neighbor}
	\psfrag{5th Relay}[Bl][Bl][1][0]{\hspace{-0em}$5^\textrm{th}$ Neighbor}
	
	\includegraphics[width=1\columnwidth]{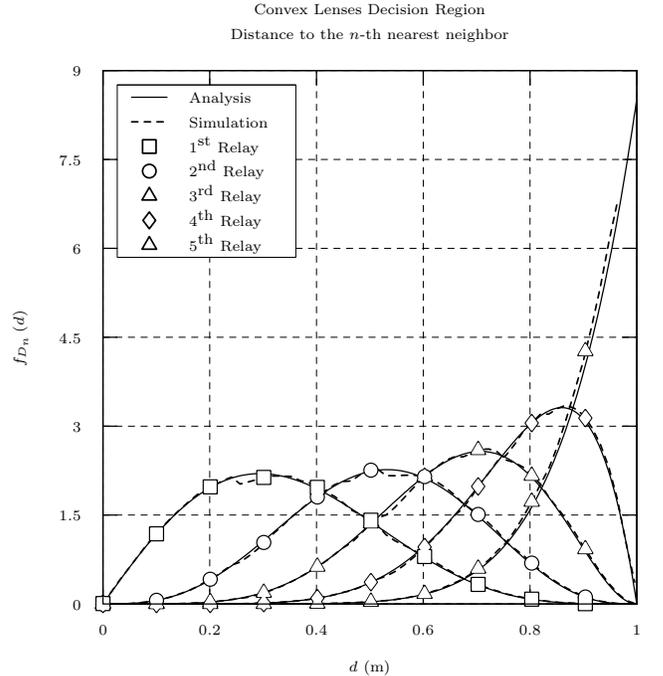}
	\vspace{-5mm}
	\caption{\ac{PDF} of the distance from a reference point
		 to the $n$-th nearest neighbor when using the \ac{CDR}
		design.  The transmission range is $R=1$m and the number of
		candidate relays  is $N=5$.}
	\label{FIG:CDR_PDF}
\end{figure}

For the $2$-dimensional Binomial processes employed here, the
\ac{CCDF} of the distance to the $n$th nearest neighbor $\bar{F}_{D_n}\left(d
\right)$ is interpreted as the probability of existing less than $n$ points
inside the decision region $\mathcal{B}$ \cite{haenggi2012stochastic}, yielding 
\vspace{-1mm}
\begin{align}
	\label{EQ:FIN_CCDF_NTH}
	\bar{F}_{D_n}\left(d \right) &=\sum_{k=0}^{n-1}{\binom {N}
	{k}p_d^k\left(1-p_d \right)^{N-k}},
\end{align}
from where \ac{PDF} $f_{D_n}(d)$ can be readily obtained.
%

\begin{figure}[!t]
	\centering
	
	\psfrag{title}[Bc][Bc][1][0]{\ac{PDF} of the distance to the nearest neighbor}
	\psfrag{ylabel}[Bc][Bc][1][0]{$f_{D_n}\left(d \right)$}
	\psfrag{xlabel}[Bc][Bc][1][0]{$d$ (m)}
	
	\psfrag{SDR}[Bl][Bl][1][0]{\hspace{-0em}SDR}
	\psfrag{CDR}[Bl][Bl][1][0]{\hspace{-0em}CDR}
	
	\psfrag{1st Round}[Bl][Bl][1][0]{\hspace{-0em}$1^\textrm{st}$ Round}
	\psfrag{2nd Round}[Bl][Bl][1][0]{\hspace{-0em}$2^\textrm{nd}$ Round}
	\psfrag{3rd Round}[Bl][Bl][1][0]{\hspace{-0em}$3^\textrm{rd}$ Round}
	
	\includegraphics[width=1\columnwidth]{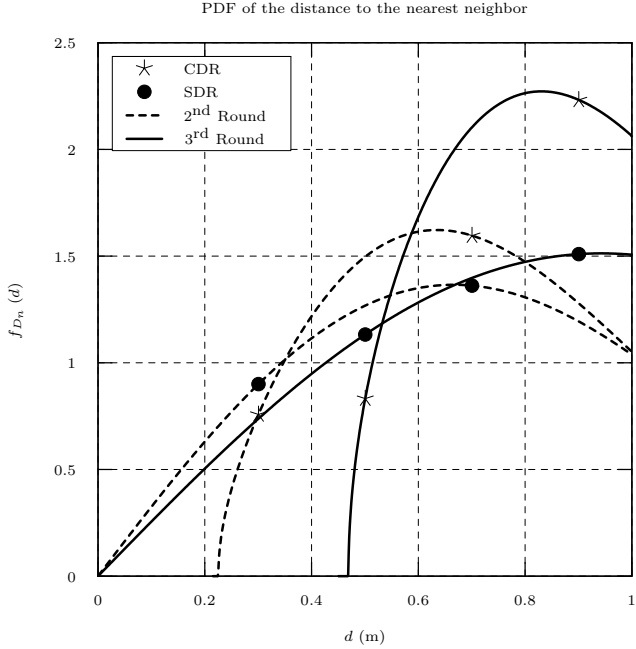}
	\vspace{-1.75mm}
	\caption{Impact of the adaptation of the decision regions on the progress regarding the nearest eligible relay. The
		transmission range is $R=1$m and the number of candidate relays  is $N=5$.}
	\label{FIG:DIST_ITER_GAIN_NEAR}
\end{figure}
\begin{figure}[!t]
	\centering
	
	\psfrag{title}[Bc][Bc][1][0]{\ac{PDF} of the distance to the furthest neighbor}
	\psfrag{ylabel}[Bc][Bc][1][0]{$f_{D_n}\left(d \right)$}
	\psfrag{xlabel}[Bc][Bc][1][0]{$d$ (m)}
	
	\psfrag{SDR}[Bl][Bl][1][0]{\hspace{-0em}SDR}
	\psfrag{CDR}[Bl][Bl][1][0]{\hspace{-0em}CDR}
	
	\psfrag{1st Round}[Bl][Bl][1][0]{\hspace{-0em}$1^\textrm{st}$ Round}
	\psfrag{2nd Round}[Bl][Bl][1][0]{\hspace{-0em}$2^\textrm{nd}$ Round}
	\psfrag{3rd Round}[Bl][Bl][1][0]{\hspace{-0em}$3^\textrm{rd}$ Round}
	
	\includegraphics[width=1\columnwidth]{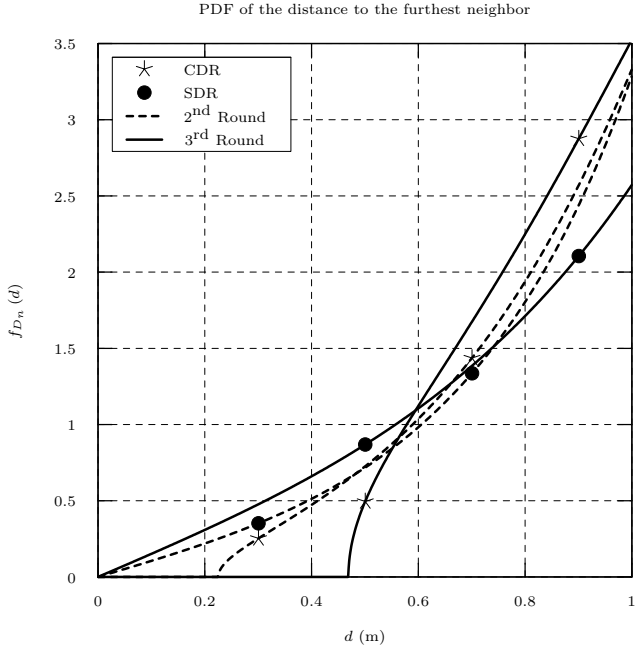}
	\vspace{-1.75mm}
	\caption{Impact of the adaptation of the decision regions on the progress regarding the furthest eligible relay. The
			transmission range is $R=1$m and the number of candidate relays  is $N=5$.}
	\label{FIG:DIST_ITER_GAIN_FAR}
\end{figure}

Fig. \ref{FIG:SDR_PDF} illustrates the distribution of the distance to the
$n$th nearest neighbor when using the \ac{SDR} design, while Fig.
\ref{FIG:CDR_PDF} presents the same distance distributions for \ac{CDR}.
Both figures were built considering the particular case of setting the radius defining the \ac{CDR} equal to the transmission range and then determining the \ac{SDR} equivalently -- by adjusting the aperture angle defining the circular sector -- in order to have the same probability of finding nodes inside the decision regions.

From these figures one can see that when there is an
equivalence between the designs of the decision regions the obtainable results are fairly comparable in terms of the achievable advancements.
When taking into consideration the effect of the recursive
iterations of the auction-based scheme on the re-computations of the forwarding regions after each collision,  the benefits of adaptively resetting the \ac{CDR} regions are evident.
Fig. \ref{FIG:DIST_ITER_GAIN_NEAR} shows the results for the nearest eligible
neighbor, while Fig. \ref{FIG:DIST_ITER_GAIN_FAR} addresses the furthest
neighbor within source's radio range.
After each collision, by considering partitions of the initial forwarding
region that are closer to the final destination, the probability of finding
nodes at further distances increases with each iteration, whereas the
likelihood of having high number of contenders in smaller areas decreases.
\vspace{-0.2cm}
\section{Results}
\label{SEC:RESULTS}

\begin{figure}[!b]
	\centering
	
	\psfrag{title}[Bc][Bc][1][0]{Expected distance to the nearest neighbor}
	\psfrag{ylabel}[Bc][Bc][1][0]{Expected separation distance (m)}
	\psfrag{xlabel}[Bc][Bc][1][0]{Initial number of contending nodes}
	
	\psfrag{1st Round}[Bl][Bl][1][0]{\hspace{-0em}CDR, $1^\textrm{st}$ Round}
	\psfrag{2nd Round}[Bl][Bl][1][0]{\hspace{-0em}CDR, $2^\textrm{nd}$ Round}
	\psfrag{3rd Round}[Bl][Bl][1][0]{\hspace{-0em}CDR, $3^\textrm{rd}$ Round}
	\psfrag{Nearest, SDR}[Bl][Bl][1][0]{\hspace{-0em}SDR}
	
	\includegraphics[width=1\columnwidth]{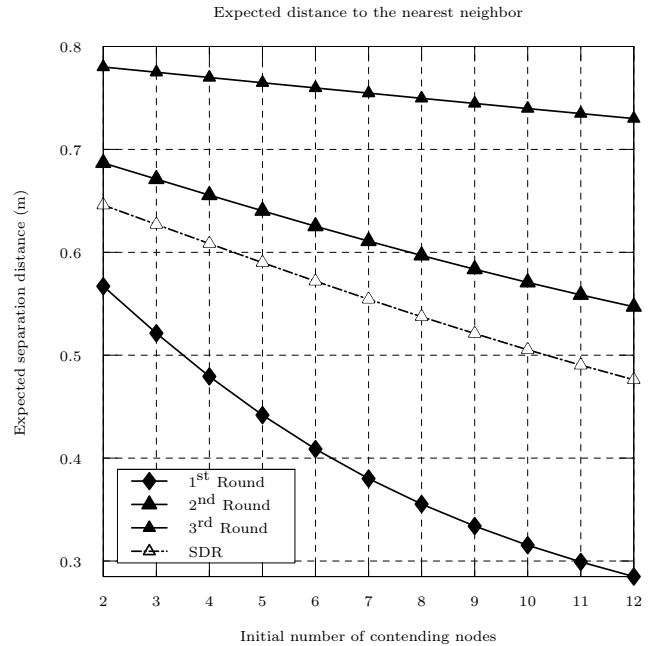}
	\vspace{-5mm}
	\caption{Expected separation distance of the nearest neighbor for the
		\ac{CGF} schemes. The source's transmission range is $R=1$m.}
	\label{FIG:EXP_NTH_NEAR}
\end{figure}

\begin{figure}[!t]
	\centering
	
	\psfrag{title}[Bc][Bc][1][0]{Expected distance to the furthest neighbor}
	\psfrag{ylabel}[Bc][Bc][1][0]{Expected separation distance (m)}
	\psfrag{xlabel}[Bc][Bc][1][0]{Initial number of contending nodes}
	
	\psfrag{1st Round}[Bl][Bl][1][0]{\hspace{-0em}CDR, $1^\textrm{st}$ Round}
	\psfrag{2nd Round}[Bl][Bl][1][0]{\hspace{-0em}CDR, $2^\textrm{nd}$ Round}
	\psfrag{3rd Round}[Bl][Bl][1][0]{\hspace{-0em}CDR, $3^\textrm{rd}$ Round}
	\psfrag{Furthest, SDR}[Bl][Bl][1][0]{\hspace{-0em}SDR}
	
	\includegraphics[width=1\columnwidth]{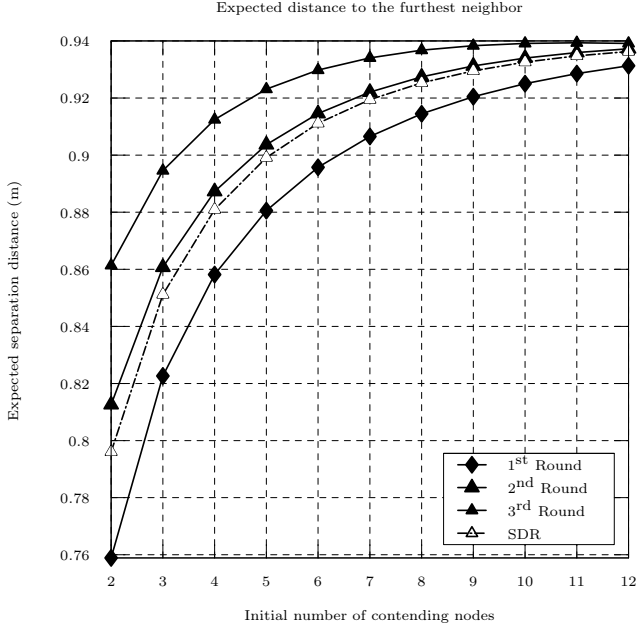}
	\vspace{-5mm}
	\caption{Expected separation distance of the furthest neighbor for the
		\ac{CGF} schemes. The source's transmission range is $R=1$m.}
	\label{FIG:EXP_NTH_FAR}
\end{figure}

We assess here the achievable progress towards the final destination and the corresponding negotiation overhead when conditioning on the initial number of colliding nodes.
Fig. \ref{FIG:EXP_NTH_NEAR} shows the expected value of the separation distance to the nearest eligible relay when using the sectoral and the convex lenses decision regions, while Fig.\ref{FIG:EXP_NTH_FAR} provides the  results for the furthest eligible relay.

Both figures show the benefit of interactively using location information.
Note that for the \ac{SDR} scheme, regardless of the previous outcome, the source node always readdresses the same original decision region for each of the subsequent rounds, while by employing the \ac{CDR} scheme nodes make use of the outputs of previous rounds for future decisions.
In this way, given that the decision regions closer to the destination enclose relays, the likelihood of hoping further increases each round.

In Fig. \ref{FIG:EXP_NTH_NEAR}, the expected distance to the nearest
relay reduces by increasing the density of nodes, but since the
\ac{CDR} is built towards the final destination and a cut-off boundary is intrinsically established, the region between source and this border is not searched for candidate relays (see Fig. \ref{FIG:CONFIG_CDR}).
Therefore, the expected distance to the nearest neighbor increases at each iteration.

\begin{figure}[!t]
	\centering
	
	\psfrag{title}[Bc][Bc][1][0]{\ac{STA}-based \ac{RSA} with \ac{SDR}}
	\psfrag{ylabel}[Bc][Bc][1][0]{Expected separation distance (m)}
	\psfrag{xlabel}[Bc][Bc][1][0]{Expected \ac{CRI} length (slots)}
	
	\psfrag{Furthest}[Bl][Bl][1][0]{\hspace{-0em}Furthest}
	\psfrag{2nd Furthest}[Bl][Bl][1][0]{\hspace{-0em}$2^\textrm{nd}$ Furthest}
	\psfrag{Nearest}[Bl][Bl][1][0]{\hspace{-0em}Nearest}
	
	\includegraphics[width=1\columnwidth]{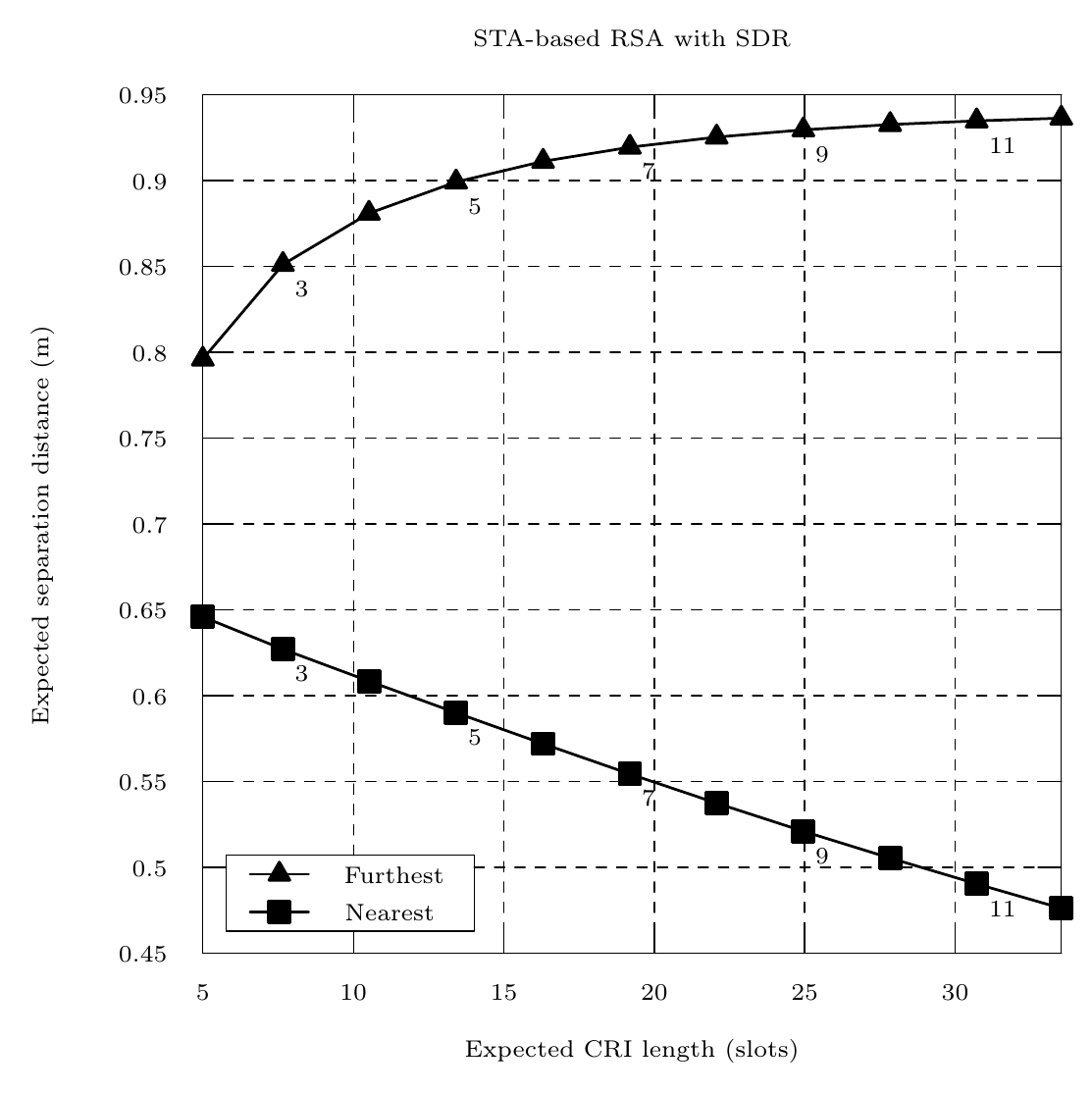}
	\vspace{-5mm}
	\caption{Expected distance to the $n$-th nearest neighbor related to the
		expected value of the \ac{CRI} length for an increasing number of
		contending relays. The numbers nearby the markers designate the initial
		number of colliding nodes. The transmission range is $R=1$m.}
	\label{FIG:EXP_NTH_NEIGHBOR_STA}
	\vspace{-3mm}
\end{figure}

For the evaluated number of relays in Fig. 
\ref{FIG:EXP_NTH_FAR}, the distance to the furthest node does not increase substantially by considering more candidates.
The more neighbors within the range, the greater is the expected distance to the furthest eligible relay.

\begin{figure}[!t]
	\centering
	
	\psfrag{title}[Bc][Bc][1][0]{Auction-based \ac{RSA} with \ac{CDR}}
	\psfrag{ylabel}[Bc][Bc][1][0]{Expected separation distance (m)}
	\psfrag{xlabel}[Bc][Bc][1][0]{Expected \ac{CRI} length (slots)}
	
	\psfrag{Furthest}[Bl][Bl][1][0]{\hspace{-0em}Furthest}
	\psfrag{2nd Furthest}[Bl][Bl][1][0]{\hspace{-0em}$2^\textrm{nd}$ Furthest}
	\psfrag{Nearest}[Bl][Bl][1][0]{\hspace{-0em}Nearest}
	
	\includegraphics[width=1\columnwidth]{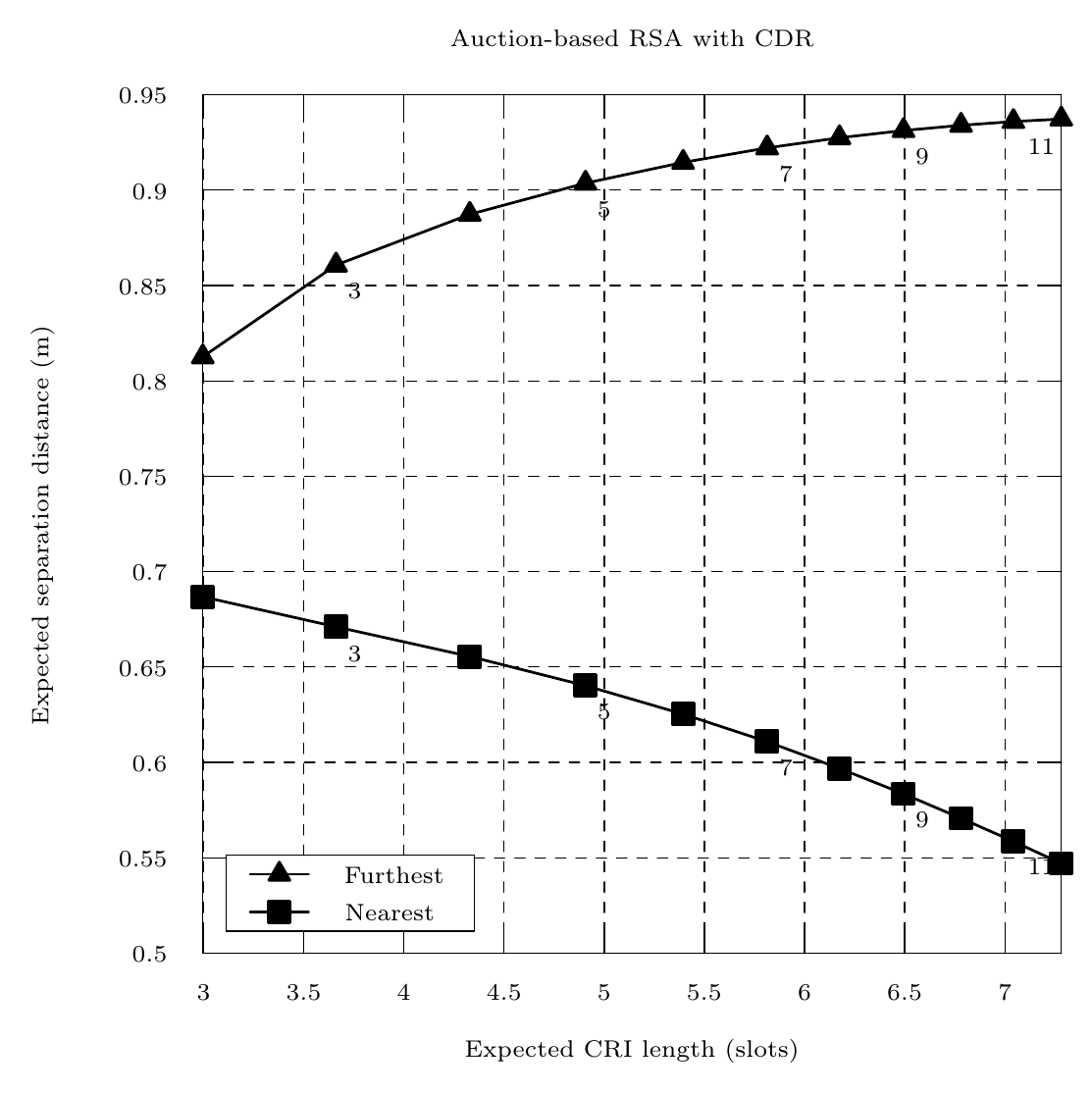}
	\vspace{-5mm}
	\caption{Expected distance to the $n$-th nearest neighbor related to the
		expected value of the \ac{CRI} length for an increasing number of
		contending relays. The numbers nearby the markers designate the initial
		number of colliding nodes. The transmission range is $R=1$m.}
	\label{FIG:EXP_NTH_NEIGHBOR_AUC}
	\vspace{-3mm}
\end{figure}

Figs. \ref{FIG:EXP_NTH_NEIGHBOR_STA} and \ref{FIG:EXP_NTH_NEIGHBOR_AUC} relate
the expected distance to the $n$-th nearest eligible relay to
the corresponding \ac{CRI} length using the \ac{STA}-based and auction-based
\ac{RSA}.
The numbers nearby the markers indicate the size of the initial
set of colliding nodes.
Regarding the expected advancement provided by the furthest neighbor strategy, the \ac{SDR} and \ac{CDR} schemes provide equivalent results.
Conversely, the expected distance of the nearest node not only experience
higher variance in hop length, but also the expected advancement become even
smaller since the nearest node is found closer to the source when the density
of candidate relays increases.

From Figs. \ref{FIG:EXP_NTH_NEIGHBOR_STA} and
\ref{FIG:EXP_NTH_NEIGHBOR_AUC}, the \ac{CRI} works against the apparent benefit of having more eligible relays within the range.
Depending on the initial number of colliding nodes, the contention resolution interactions may linger too long and then compromise the performance.
It is then reasonable to maintain small the number of potential relays that get actively involved in the election process, because the selection procedure substantially contributes to the communication cost at hop-basis.
%
%

\section{Final Remarks}
\label{SEC:CONCLUSIONS}

In this paper, geographic routing strategies are studied by
assessing their constituent operational parts.
The \ac{CRI} length required to find a relay in multi-hop scenarios is characterized and the progress that is enabled by the forwarding decision regions are described.
Our results show that the proposed auction-based approach using location information outperforms the \ac{STA}-based solution.
The auction-based \ac{RSA} substantially reduces the protocol overhead of establishing active connections in autonomous multi-hop networks allowing for more efficient reuse of the shared channel.

\end{document}